\documentclass[lettersize,journal]{IEEEtran}
\usepackage[utf8]{inputenc}
\usepackage[T1]{fontenc}
\usepackage[english]{babel}

\usepackage{url}
\usepackage[numbers]{natbib}
\usepackage{graphicx} 
\usepackage{booktabs}
\usepackage{balance}
\usepackage{multirow}
\usepackage{colortbl}
\usepackage{listings}

\usepackage{xcolor}
\usepackage{enumitem}
\usepackage{amsmath,amssymb}
\usepackage{xspace}

\newcommand{\ie}{\textit{i.e.,}\xspace}
\newcommand{\eg}{\textit{e.g.,}\xspace}
\newcommand{\darkred}{\color[RGB]{139,0,0}}
\newcommand{\darkgreen}{\color[RGB]{0,100,0}}
\definecolor{darkgreen}{rgb}{0.0, 0.5, 0.0}

\usepackage[skins]{tcolorbox}
\usepackage{listings}

\usepackage{tcolorbox}

\usepackage{fontawesome5}

\begin{document}
\title{Smaller = Weaker? Benchmarking Robustness of \\ Quantized LLMs in Code Generation}

\author{Sen Fang, Weiyuan Ding, Antonio Mastropaolo and~Bowen Xu
\IEEEcompsocitemizethanks{\IEEEcompsocthanksitem S. Fang, W. Ding, and B. Xu are with NC State University, USA. Antonio Mastropaolo is with William \& Mary, USA.}
\thanks{E-mail: \{sfang9, wding8, bxu22\}@ncsu.edu, amastropaolo@wm.edu}
}

\maketitle

\begin{abstract}
Quantization has emerged as a mainstream method for compressing Large Language Models (LLMs), reducing memory requirements and accelerating inference without architectural modifications.
While existing research primarily focuses on evaluating the effectiveness of quantized LLMs compared to their original counterparts, the impact on robustness remains largely unexplored.
In this paper, we present the first systematic investigation of how quantization affects the robustness of LLMs in code generation tasks. Through extensive experiments across four prominent LLM families (LLaMA, DeepSeek, CodeGen, and StarCoder) with parameter scales ranging from 350M to 33B, we evaluate robustness from dual perspectives: adversarial attacks on input prompts and noise perturbations on model architecture.
Our findings challenge conventional wisdom by demonstrating that quantized LLMs often exhibit superior robustness compared to their full-precision counterparts, with 51.59\% versus 42.86\% of our adversarial experiments showing better resilience in quantized LLMs. 
Similarly, our noise perturbation experiments also confirm that LLMs after quantitation generally withstand higher levels of weight disturbances. 
These results suggest that quantization not only reduces computational requirements but can actually enhance LLMs' reliability in code generation tasks, providing valuable insights for developing more robust and efficient LLM deployment strategies.
\end{abstract}

\section{Introduction}

Quantization~\cite{krishnamoorthi2018quantizing, liang2021pruning, gholami2022survey} has gained significant attention as an effective method for compressing large language models (LLMs). This technique is advantageous not only for general-purpose LLMs but also for optimizing models designed for specialized tasks, such as large code models (LCMs)\footnote{In this paper, we use LLMs and LCMs interchangeably}. In particular, LCMs employ distinct optimization strategies to enhance the automation of software engineering (SE) tasks such as bug fixing, source code generation and code review, among others \cite{watson2022systematic,hou2024large}.

The increasing demand for compressed LLMs has led to a surge in publicly available quantized models, many of which are hosted on platforms like Hugging Face. Notably, nearly all state-of-the-art LLMs now have quantized versions, enabling more efficient deployment without significant performance loss. Examples include DeepSeek~\cite{deepseek} and LLaMA~\cite{llama}, both of which have been successfully quantized and made publicly accessible for broader use. 

Unlike knowledge distillation~\cite{hinton2015distilling, jiang2023knod}, another model compression technique, which requires extensive retraining, quantization directly converts high-precision weights to lower-precision representations, reducing memory footprint and accelerating inference without architectural modifications. 
Moreover, it leverages widespread hardware support for low-bit operations, enabling efficient deployment on edge devices~\cite{liu2021post}. Its non-parametric nature preserves model structure while offering flexible precision-performance trade-offs through various granularities and bit widths, making it particularly effective for LLMs~\cite{dettmers2023qlora, egashira2024exploiting}.

Existing studies pay more attention to evaluate the effectiveness of quantized LLMs compared to their original versions. For example, \cite{afrin2025resource} compared the effectiveness of the original and quantized model in code summarization task. \cite{weyssow2023exploring} studied multiple parameter-efficient fine-tuning techniques and verified that LLMs fine-tuned with QLoRA~\cite{dettmers2023qlora} could achieve competitive performance compared with full fine-tuning.
Overall, many existing works~\cite{wei2023towards, imani2024context, egashira2024exploiting, lin2024awq, zhao2024atom} demonstrate that the quantized LLMs can achieve performance comparable to their original versions in terms of effectiveness.

\begin{figure}[th]
\begin{center}

\begin{lstlisting}[
frame=single, 
breaklines=true, 
breakindent=0pt, 
columns=fullflexible, 
mathescape=true,
basicstyle=\footnotesize,
keywords={given, specified},
keywordstyle=\color{red},
caption={The prompt from MBPP$^+$ No. 792 task. We perform sentence level adversarial attack by translating the prompt into German then translating back.}, 
captionpos=b, 
label={lst:motivation-exp}
]
$\textbf{Original prompt:}$ 
"""
Write a python function to count the number of lists in a given number of lists.
assert count_list([[1, 3], [5, 7], [9, 11], [13, 15, 17]]) == 4
"""
$\textbf{Adversarial prompt:}$
"""
Write a python function to count the number of lists in a specified number of lists.
assert count_list([[1, 3], [5, 7], [9, 11], [13, 15, 17]]) == 4
"""
\end{lstlisting}
\begin{minipage}[t]{.18\textwidth}
\begin{lstlisting}[language=Python, columns=flexible, frame=single, basicstyle=\footnotesize, label={lst:ori_solution}, caption={Output of original model with original prompt. \faThumbsUp}, captionpos=b, breaklines=true, keywordstyle=\color{blue}, commentstyle=\color{dkgreen}, stringstyle=\color{mauve}]
def count_list(lst):
  count = 0
  for i in lst:
    if isinstance(i, list):
      count += 1
  return count
\end{lstlisting}
\end{minipage}
\hfill
\begin{minipage}[t]{.18\textwidth}
\begin{lstlisting}[language=Python, columns=flexible, frame=single, basicstyle=\footnotesize, label={lst:adv_solution}, caption={Output of original model with adversarial prompt. \faThumbsDown}, captionpos=b, breaklines=true, keywordstyle=\color{blue}, commentstyle=\color{dkgreen}, stringstyle=\color{mauve}]
def count_list(lst):
  return len(lst)
\end{lstlisting}
\end{minipage}
\hfill
\begin{minipage}[t]{.18\textwidth}
\begin{lstlisting}[language=Python, columns=flexible, frame=single, basicstyle=\footnotesize, label={lst:qori_solution}, caption={Output of quantized model with original prompt. \faThumbsUp}, captionpos=b, breaklines=true, keywordstyle=\color{blue}, commentstyle=\color{dkgreen}, stringstyle=\color{mauve}]
def count_list(lst):
  count = 0
  for i in lst:
    if type(i) == list:
      count += 1
  return count
\end{lstlisting}
\end{minipage}
\hfill
\begin{minipage}[t]{.18\textwidth}
\begin{lstlisting}[language=Python, columns=flexible, frame=single, basicstyle=\footnotesize, label={lst:qadv_solution}, caption={Output of quantized model with adversarial prompt. \faThumbsUp}, captionpos=b, breaklines=true, keywordstyle=\color{blue}, commentstyle=\color{dkgreen}, stringstyle=\color{mauve}]
def count_list(lst):
  count = 0
    for i in lst:
      if type(i) == list:
        count += 1
    return count
\end{lstlisting}
\end{minipage}
\end{center}
\caption{The generated solutions for MBPP$^+$ No. 792 task by DeepSeek-Coder-6.7B pre- and post-quantization with prompts before and after attack.}
\label{fig:motivation-exp}
\end{figure}
However, from the analysis we conducted, a question naturally emerges: \textit{Does ``free lunch'' really exist?} 
We look into this question through the lens of another critical property of LLMs: \ie robustness.
As illustrated in the Figure~\ref{lst:motivation-exp}, we apply a sentence-level adversarial attack~\cite{wei2023towards} to generate an adversarial prompt for the code generation task. 
The adversarial attack translates the original English prompt into German and then back to English, creating a semantically equivalent prompt with minor differences—specifically, replacing the word ``given'' with ``specified''.
When we presented both prompts to DeepSeek-Coder-6.7B-base~\footnote{https://huggingface.co/deepseek-ai/deepseek-coder-6.7b-base}, we observed an intriguing phenomenon. While the model pre- and post-quantization both correctly handle the original prompt, their behaviors diverge significantly on the adversarial prompt. The original model misinterprets the adversarial prompt entirely, generating a solution that counts the length of a given list rather than counting the number of list elements in the given list. In contrast, the quantized model maintains consistent and correct behavior across both prompts, generating identical solutions that properly count the number of lists. Moreover, same situation is also observed in LLaMA models.
Our observation is counter-intuitive is because, quantization is typically viewed as a compression technique that trades off model performance for efficiency, actually enhances model robustness – raising an intriguing question about the relationship between quantization and LLMs' behavior.

Hence, motivated by our observation, we aim to answer the following key question,
\begin{tcolorbox}
\textit{How does quantization impact LLMs' robustness in code generation task?}
\end{tcolorbox}

We specifically focus on code generation for several compelling reasons~\cite{liang2024large}. First, code generation serves as a rigorous benchmark, requiring both precise logical reasoning and strict syntactic correctness—even minor errors can render the output completely non-functional. Second, the widespread adoption of AI-assisted programming tools~\cite{liang2024large, davila2024industry} in software development workflows has made code generation one of the most commercially significant applications of LLMs. This practical importance is reflected in the fact that coding ability has become a standard evaluation criterion for newly released LLMs~\cite{zhu2024deepseek, dubey2024llama}, which are routinely assessed on important benchmarks like HumanEval~\cite{chen2021evaluating} and MBPP~\cite{austin2021program}. Third, the robustness of code generators has already been a thread of research~\cite{mastropaolo2023robustness, wei2023towards, jha2023codeattack, qu2024survey, zhang2024attacks, zhang2023transfer}, highlighting the significance of this challenge and its broader implications. Moreover, code generation presents distinct challenges for model quantization, as it requires maintaining accuracy across both natural language understanding and formal programming language constraints, effectively providing a stringent test of how quantization affects LLMs' robustness.

To answer the above question, we conduct a systematic investigation to benchmark the robustness of LLMs pre- and post-quantization from two different perspectives: adversarial attack on the input prompt and noise perturbation on the model architecture.
We consider three different adversarial attacks from different levels, as well as two mainstream noise perturbation methods, performing experiments on four popular LLM families (\ie LLaMA~\cite{dubey2024llama}, DeepSeek~\cite{zhu2024deepseek}, CodeGen~\cite{nijkamp2023codegen}, and StarCoder~\cite{lozhkov2024starcoder}) with scales from 350M to 33B.

Surprisingly, our results reveal that quantized LLMs often exhibit greater robustness than their original versions, both in handling adversarial inputs and withstanding architectural noise. Specifically, in our extensive adversarial experiments, quantized LLMs successfully defended against more adversarial attacks in a significant majority of cases. Similar patterns emerged in our noise perturbation experiments, where quantized LLMs generally withstood a broader spectrum of noise attacks from various sources, proving that the common sense that \emph{``smaller=weaker''} may not be the case in the context of quantization LLM for code generation.

In summary, we make the following contributions:

\begin{itemize}
    \item We present the first comprehensive investigation exploring how quantization affects the robustness of LLMs in code generation tasks. Our work challenges the conventional wisdom that quantization necessarily trades model quality for efficiency.
    \item We evaluate robustness through dual complementary perspectives: input-level robustness using three types of adversarial attacks and model-level robustness through two systematic noise perturbations.
    \item Our extensive experiments across four prominent LLM families (LLaMA, DeepSeek, CodeGen, and StarCoder) with parameter scales ranging from 350M to 33B provide compelling evidence that quantized LLMs often exhibit superior robustness compared to their full-precision counterparts.
    \item We develop and open-source a flexible, standardized tool~\footnote{https://github.com/safeai4code/adversarial-codegen} for evaluating LLM robustness pre- and post-quantization, which is applicable to any LLM, facilitating future research.
\end{itemize}

\section{Related Works}

\subsection{LLMs' Robustness in Code Generation}

Code generation has been identified as the most popular AI application among various coding tasks~\cite{liang2024large}. Hence, recent studies have extensively investigated the robustness of LLMs in code generation \cite{wei2023towards, jha2023codeattack, mastropaolo2023robustness, shirafuji2023exploring, zhuo2023robustness, chen2023evaluating, zhang2023transfer, yang2024robustness, ge2024demonstration, liu2024alanca, zhang2024attacks, li2024attribution, improta2025enhancing}. \citeauthor{mastropaolo2023robustness} empirically evaluated the robustness of GitHub Copilot by submitting semantically equivalent natural language description pairs. Their results indicated that Copilot produced different code recommendations in 46\% of cases, with a corresponding 28\% gap in correctness between these variants. Similarly, \citeauthor{shirafuji2023exploring} and \citeauthor{zhuo2023robustness} both proved that Codex is highly sensitive to even minor modifications in the natural language description~\cite{shirafuji2023exploring, zhuo2023robustness}. Moreover, \citeauthor{chen2023evaluating, zhang2023transfer, zhang2024attacks}, and \citeauthor{improta2025enhancing} have shown that LLMs are vulnerable to adversarial attacks from various perspectives and have proposed different approaches to enhance their robustness~\cite{chen2023evaluating, zhang2023transfer, zhang2024attacks, improta2025enhancing}. Notably, \citeauthor{wei2023towards} employed three distinct adversarial attacks to assess the robustness of LLMs pre- and post-quantization, finding that quantized models maintain robustness levels comparable to their original versions~\cite{wei2023towards}. Besides, \citeauthor{qu2024survey} identified critical research directions for enhancing LLMs' robustness in the code generation task, including developing realistic attack scenarios, improving risk assessment frameworks, and exploring to combine formal verification with neural networks to provide theoretical guarantees of robustness~\cite{qu2024survey}.

Our work advances the field in three key aspects. First, we conduct a comprehensive robustness evaluation of quantized LLMs from dual perspectives: input-level robustness through various adversarial attack methods, and model-level robustness through parameter weight perturbations. Second, through these complementary analyses, we demonstrate that quantized LLMs can achieve superior robustness compared to their original counterparts. Third, we present a detailed quantitative analysis explaining the mechanisms behind the enhanced robustness of quantized LLMs.

\subsection{Quantized LLMs for Software Engineering}

While quantization has been widely adopted in general-purpose LLMs, its application in software engineering LLMs remains limited \cite{wei2023towards, kaushal2023lord}. \citeauthor{wei2023towards} conducted pioneering empirical studies on quantizing LLMs for code generation, demonstrating that 8-bit quantization can reduce model size and inference latency while maintaining competitive performance~\cite{wei2023towards}. Their experiments across multiple programming languages showed a marginal 1.2\% decrease in code generation correctness. In a parallel effort, \citeauthor{kaushal2023lord} proposed a low-rank decomposition method combined with quantization for LLMs, achieving up to a 4× compression ratio with minimal impact on model effectiveness~\cite{kaushal2023lord}. \citeauthor{weyssow2023exploring} performed a comprehensive study on using QLoRA for code generation under resource constraints, demonstrating that QLoRA can reduce memory usage by up to 2× compared to standard LoRA while maintaining or improving model effectiveness~\cite{weyssow2023exploring}. Their experiments showed that QLoRA enabled fine-tuning of large models up to 34B parameters using less than 24GB of GPU memory. For example, QLoRA-4bit achieved a remarkable 12.2\% increase in performance over CodeLlama-7B-Python with LoRA on the Conala dataset, while CodeLlama-34B-Python fine-tuned with QLoRA-4bit achieved state-of-the-art performance on both Conala and CodeAlpacaPy benchmarks.
More recently, \citeauthor{afrin2025resource} explored QLoRA for code summarization tasks, and their comprehensive study on CodeLlama and DeepSeek-Coder models showed that QLoRA not only matches but consistently outperforms full model fine-tuning, delivering superior results while significantly improving resource efficiency~\cite{afrin2025resource}. Specifically, QLoRA achieved a 2–3\% improvement in METEOR scores across Python and Java tasks while reducing the memory footprint by approximately one-third. Their qualitative analysis further revealed that QLoRA-optimized models can generate more accurate and informative code summaries than those written by developers in some specific cases.

Overall, our pioneering work is the first to systematically demonstrate that LLMs with 8-bit quantization not only retain but can enhance robustness compared to their full-precision counterparts in code generation tasks. 
Our findings strongly support the adoption of quantized LLMs in software engineering, not only to enhance model sustainability but also to strengthen the robustness of pipelines that integrate code generators as AI-powered development tools.

\section{Methodology}

\begin{figure*}[t]
    \centering
    \includegraphics[width=1\linewidth]{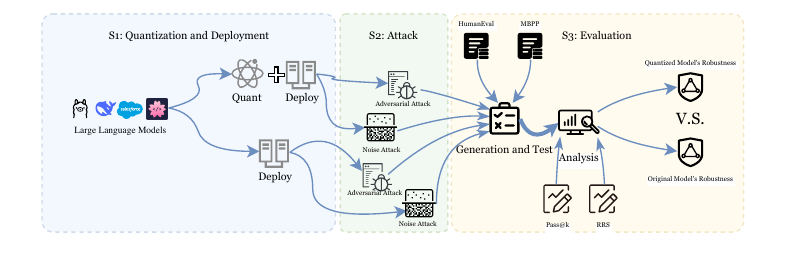}
    \caption{The overview of our methodology.}
    \label{fig:method}
\end{figure*}
Our work consists of three fundamental stages: (S1) quantization; (S2) robustness attacks and; (S3) evaluation, as illustrated in Figure~\ref{fig:method}. The first stage focuses on the selection of target LLMs and appropriate quantization methods. The second stage applies selected attack strategies on LLMs both pre- and post-quantization, approaching from two distinct angles: adversarial attacks targeting the input, and noise attacks targeting the model architecture. In the final stage, we conduct a comprehensive evaluation of how quantization influences LLMs robustness by analyzing the outputs of each model under various attack conditions.

\subsection{S1: Selection of Quantization Method}
Generally, quantization method could be categoried into the following two classes:
\begin{itemize}[wide, labelwidth=!, labelindent=0pt]

    \item \textbf{Dynamic Quantization}~\cite{kim2023full, nagel2021white} works at the inference phase, where scaling factors are computed dynamically based on the actual tensor values in each forward pass. For LLMs, this quantization strategy is particularly suitable for attention layers and feed-forward networks, where activation patterns can vary significantly across different inputs. Dynamic quantization offers flexibility in handling varying input distributions but may introduce additional computational overhead during inference. Most implementations focus on quantizing the weights and activations to INT8 or lower precision formats while keeping critical components like layer normalization in higher precision to maintain stability.

    \item \textbf{Static Quantization} \cite{shen2020q, frantar2022gptq, dettmers2022gpt3, lin2024awq} determines quantization parameters prior to deployment, all of which remain fixed during inference. Currently, several advanced static quantization methods have been developed specially for LLMs. For instance, post-training quantization schemes like NF4 and FP4 use predefined parameters optimized for transformer architectures, eliminating the need for calibration data required in traditional techniques. In other hand, vector-wise quantization approaches such as \texttt{GPTQ} analyze entire weight matrices to preserve geometric properties. Generally, these specified quantization methods typically achieve better performance than dynamic quantization due to their optimization for transformer structures, reducing runtime overhead.
    
\end{itemize}


In this work, we select \texttt{bitsandbytes}\footnote{https://github.com/bitsandbytes-foundation/bitsandbytes} to quantize LLMs in the static way. There are some reasons~\cite{dettmers20218, hf_bitsandbytes}: First, it implements specialized quantization schemes (NF4/FP4) that are specifically optimized for transformer architectures, maintaining the complex patterns and geometric properties learned during pre-training while achieving significant memory reduction. Second, it offers exceptional ease of integration within the modern LLMs development ecosystem, with seamless compatibility with the Hugging Face transformers\footnote{https://huggingface.co/docs/transformersd}, thus it has been extensively used in LLMs quantization. Third, it provides a robust balance between memory efficiency and computational performance, achieving up to 4x memory reduction compared to FP16 models while maintaining competitive inference speed. Additionally, \texttt{bitsandbytes} eliminates the need for task-specific calibration data, which is particularly advantageous when working with LLMs where obtaining representative calibration datasets can be challenging. It also supports mixed-precision operations, allowing critical layers to remain in higher precision while aggressively quantizing less sensitive components, ensuring that accuracy is preserved for key model functionalities.

\subsection{S2: Robustness Attack Methods}


We attack LLMs pre- and post-quantization from two different perspectives, \ie adversarial attacks on input and noise attacks on model architecture. For adversarial attacks, we employ three widely-used methods that have been extensively established in prior work \cite{wei2023towards, jha2023codeattack,mastropaolo2023robustness, improta2025enhancing}:

\begin{itemize}[wide, labelwidth=!, labelindent=0pt]

    \item \textbf{Character-level}: 
        This method randomly converts selected characters in the input prompt to uppercase letters. While preserving the semantic meaning, this simple perturbation tests the model's sensitivity to character case variations.

    \item \textbf{Word-level}:
        This method randomly substitutes words in the prompt with their WordNet~\cite{miller1995wordnet} synonyms. It assesses whether the model maintains consistent performance when encountering semantically equivalent alternatives while preserving the original prompt's meaning.

    \item \textbf{Sentence-level}:
        This method employs back-translation, converting the English prompt to German and then back to English using a task-specific LLM. The resulting paraphrased versions maintain the core meaning while testing the model's ability to handle semantic-preserving reformulations.
        
\end{itemize}

For noise attacks, we employ two methods that have been extensively used to perturb LLMs \cite{liu2018analyzing,gong2024makes,galli2024noisy}: 

\begin{itemize}[wide, labelwidth=!, labelindent=0pt]
\item \textbf{Gaussian noise}:
Add random perturbations drawn from a normal distribution ($\mathcal{N}(\mu, \sigma^2)$) to the model weights. It follows a bell-shaped distribution and is widely used in statistical analysis, simulating natural random variations in the weight parameters while maintaining specific statistical properties.
\item \textbf{Uniform noise}:
Apply random perturbations sampled from a uniform distribution ($\mathcal{U}(a, b)$) to the model weights. It provides equal probability for all values within a specified range, evaluating the model's resilience to consistent magnitude perturbations across all weight parameters.
\end{itemize}

\subsection{S3: Evaluation}

\noindent\textbf{Task effectiveness measurement.}
For each programming problem in our dataset, a dedicated test suite is provided to assess the accuracy of the generated solutions. By following prior works on code generation~\cite{wei2023towards, zhu2024deepseek}, we reuse the same metric to evaluate a model’s performance in code generation, \ie  if a generated program successfully passes all tests in the suite, it is considered a pass. In addition, the \texttt{pass@k} metric quantifies the likelihood of finding a working solution among $k$ generated samples. Specifically, for each problem, the model generates $k$ samples, and \texttt{pass@k} is defined as the probability that at least one of these samples is correct.

\noindent\textbf{Robustness measurement.} Previous work \cite{wei2023towards} have evaluated the robustness of LLMs pre- and post-quantization by calculating the percentage change in performance between perturbed and unperturbed inputs (as shown in Equation~\ref{eq:delta_percentage}),


\begin{equation}
\label{eq:delta_percentage}
\%\Delta = \frac{\text{pass@1}_{\text{clean}} - \text{pass@1}_{\text{attack}}}{\text{pass@1}_{\text{clean}}}
\end{equation}

However, it has several limitations when analyzing the impact of quantization on LLMs' robustness:  (1) it requires separate calculations rather than providing a direct comparative measure;  (2) although it calculates relative performance differences within each model, the different baseline performance levels between original and quantized models is overlooked; (3) it lacks a standardized threshold to determine whether quantization improves or degrades relative robustness across models; and (4) it does not provide a normalized comparison of robustness degradation.

Inspired by previous work that focus on normalized comparison~\cite{hendrycks2019benchmarking, tramer2020adaptive, subbaswamy2021evaluating, zhang2020attacks, hendrycks2016baseline}, we propose the Relative Robustness Score (RRS), which directly quantifies the relationship between robustness degradation in original and quantized models:
\begin{equation}
\text{RRS} = \frac{pass@1_{\text{orig}}^{\text{clean}} - pass@1_{\text{orig}}^{\text{attack}}}{pass@1_{\text{quant}}^{\text{clean}} - pass@1_{\text{quant}}^{\text{attack}}},
\end{equation}
Here, $pass@1_{\text{orig}}^{\text{clean}}$ and $pass@1_{\text{quant}}^{\text{clean}}$ denote the \texttt{pass@1} scores of the original and quantized models under normal (unperturbed) conditions, respectively, while $pass@1_{\text{orig}}^{\text{attack}}$ and $pass@1_{\text{quant}}^{\text{attack}}$ represent their corresponding scores under perturbations. An RRS value greater than 1 indicates that the quantized model experiences less performance degradation than the original model, thereby demonstrating better robustness against perturbations. 

RRS offers several advantages: First, it provides a direct comparative measure between original and quantized models. Second, it establishes a clear interpretative threshold where RRS > 1 indicates enhanced robustness in the quantized model, facilitating straightforward assessment of quantization effects. Third, by normalizing performance degradation across both model variants, RRS accounts for baseline performance differences, ensuring fair comparisons even when original and quantized models exhibit different accuracy levels under clean conditions.

\section{Research Questions \& Settings}

\subsection{Research Questions}

In this work, we focus on the following research questions:
\begin{itemize}[wide, labelwidth=!, labelindent=0pt]

    \item \textbf{RQ1 (Adversarial Robustness Evaluation)}:
    How does quantization affect the robustness of LLMs against existing adversarial attacks in code generation?

    \item \textbf{RQ2 (Noise Perturbation Evaluation)}:
    How does quantization affect the robustness of LLMs against noise attack on the weights of LLMs in code generation?

    \item \textbf{RQ3 (Quantization-Robustness Trade-off Evaluation)}:
    How does quantization create a trade-off between performance and resilience for LLMs in code generation task?

\end{itemize}

\subsection{Studied Models}
We select LLMs by prioritizing models that have been extensively studied and validated for their performance in code generation tasks \cite{deroy2024code,roziere2023code,zhu2024deepseek,yu2024codereval,feng2024complexcodeeval,zeng2024coderujb,zheng2024towards}.
As a result, our selection includes the following models:


\begin{itemize}[wide, labelwidth=!, labelindent=0pt]

    \item \textbf{LLaMA Family}: 
    LLaMA \cite{dubey2024llama}, developed by Meta AI, is a family of open-source LLMs that have achieved state-of-the-art performance across numerous evaluation benchmarks. In our experiments, we evaluate three variants: LLaMA-3.2-1B, LLaMA-3.2-3B, and LLaMA-3.1-8B. 
    
    \item \textbf{DeepSeek Family}:
    DeepSeek \cite{zhu2024deepseek}, another state-of-the-art LLM series developed by DeepSeek AI, has demonstrated superior performance across diverse NLP and SE tasks. In this work, we utilize three code-specific variants: DeepSeek-Coder-1.3B-base, DeepSeek-Coder-6.7B-base, and DeepSeek-Coder-33B-base.

    \item \textbf{CodeGen Family}:
    CodeGen \cite{nijkamp2023codegen}, developed by Salesforce Research, is a family of LLMs specifically trained for code. They are trained on a large corpus of codebase from various programming languages and have shown strong performance in code completion and generation tasks. In this work, we include three variants: CodeGen-350M, CodeGen-2B, and CodeGen-6B, the mono version.
    
    \item \textbf{StarCoder Family}:
    StarCoder \cite{lozhkov2024starcoder}, a notable advancement in LCMs, demonstrates exceptional performance in code understanding and generation tasks across multiple programming languages. In our study, we evaluate three model variants - StarCoder2-3B, StarCoder2-7B, and StarCoder2-15B.

\end{itemize}

\subsection{Datasets}
We evaluate our approach on two mainstream code generation benchmarks: HumanEval \cite{chen2021evaluating} and MBPP \cite{austin2021program}. HumanEval consists of 164 Python programming problems with test cases, including string manipulation, mathematical operations, and data structure algorithms. MBPP (Mostly Basic Programming Problems) contains 974 Python programming tasks, designed to test fundamental programming abilities. To enhance result reliability, we additionally incorporate HumanEval Plus and MBPP Plus\footnote{MBPP Plus filters low-quality samples to 378 problems, so we evaluate identical samples across both versions.} \cite{liu2024your}, which expand the original test suites by 80x and 35x respectively, providing more comprehensive validation of model performance.

\subsection{Model Quantization}

We leverage \texttt{bitsandbytes} library\footnote{\url{https://huggingface.co/docs/bitsandbytes}} for model quantization, which supports both 4-bit and 8-bit precision levels. For 4-bit quantization, we employed the NF4 (normalized float 4) format with double quantization enabled and float32 compute dtype to balance efficiency with numerical stability. The 8-bit quantization uses standard linear quantization applied uniformly across model weights, all of which are built through the Hugging Face transformers library's BitsAndBytesConfig. With this quantization method, we could maintain consistent tokenization and generation capabilities between original and quantized models, ensuring fair comparison when evaluating robustness.

\subsection{Robustness Attacks}


\noindent\textit{\textbf{Adversarial Attack Implementation}} We implement three distinct adversarial attack methods in different levels:
\begin{itemize}[wide, labelwidth=!, labelindent=0pt]

    \item \textbf{Character-level Attack}: 
    We implemented by identifying all alphabetic characters in the prompt and randomly selecting a subset for transformation to uppercase. The attack uses a configurable character change probability (set to 0.5 in our experiments) and enforces a maximum change limit (5 characters) to control perturbation intensity. The implementation first identifies all eligible alphabetic positions, calculates the probable number of changes based on the probability parameter, and then applies random sampling to select which characters to modify.

    \item \textbf{Word-level Attack}: 
    In our implementation, we replace content words with synonyms from WordNet based on their part-of-speech tags. The attack targets specific POS categories (nouns, verbs, adjectives, adverbs) while excluding stopwords to preserve meaning. For each replaceable word, the system retrieves up to a maximum number of synonyms (3 in our experiments) and randomly selects one, with each replacement occurring with a probability of 0.15. We maintain proper text reconstruction by handling punctuation spacing patterns.

    \item \textbf{Sentence-level Attack}: 
    We implemented back-translation using the Facebook \texttt{mbart-large-50-many-to-many-mmt}\footnote{https://huggingface.co/facebook/mbart-large-50-many-to-many-mmt} model in a two-step pipeline. The attack first translates English text to German using the ``en\_XX'' to ``de\_DE'' configuration, then translates the result back to English using ``de\_DE'' to ``en\_XX'' setting. This creates natural paraphrases with subtly different word choices and sentence structures while maintaining semantic equivalence. We build it by leveraing Hugging Face transformers pipeline API.
    
\end{itemize}

All attacks are designed to only perturb the natural language portion of prompts by preserving code structure and test case. We set a consistent random seed to ensure reproducibility.

\noindent\textit{\textbf{Noise Perturbation Implementation}} We evaluate architectural robustness by implementing controlled parameter noise perturbations across multiple intensity levels. By using PyTorch's native functions, we produce two distinct noise sources: Gaussian noise via $torch.randn\_like()$ for normal distribution perturbations;  and Uniform noise using $torch.randn\_like()$ transformed to balanced [-noise\_level, +noise\_level] ranges through $(torch.rand\_like(param) * 2 - 1) * noise\_level$. We systematically test each noise type at five increasing intensities $(1e-4, 1e-3, 3e-3, 5e-3, and, 1e-2)$, providing a granular assessment across a 100x range.

We optimizes efficiency by wrapping all operations in PyTorch's $no\_grad()$ context, selectively targeting only gradient-related parameters, and applying perturbations through in-place $add\_()$ operations to minimize memory overhead. This methodical approach enables precise identification of critical thresholds where model performance degrades, yielding insights into how quantization affects robustness against parameter perturbations.

\subsection{Evaluation Framework}

Our evaluation framework assesses LLMs robustness through a two-step process. We leverage EvalPlus to evaluate the functional correctness of generated solutions against comprehensive test suites. It executes generated code in a secure sandbox environment, comparing outputs against ground-truth references with appropriate tolerance parameters. For each model variant and perturbation technique, we calculate the pass@1 metric to measure base performance. To quantify robustness differences between original and quantized models, we implemented a custom Python script to calculate our proposed Relative Robustness Score (RRS). This score directly compares performance degradation ratios between original and quantized models under identical perturbation conditions.

\subsection{Experimental Environment}
We conducted all experiments on a Ubuntu server with an AMD Ryzen Threadripper PRO 7955WX CPU, 256 GB RAM, and dual NVIDIA A6000 GPUs. We use Hugging Face \texttt{transformers} to use LLMs under different attack settings for code generation tasks.

\section{Results}
\begin{table*}[t]
\centering
\caption{Robustness comparison of original and quantized LLMs under different adversarial attacks. Ch, W, and S represent \textbf{CH}aracter-level, \textbf{W}ord-level, and \textbf{S}entence-level attacks, respectively. $\darkgreen\Uparrow$ means original LLMs are more robust, and $\darkred\Downarrow$ means quantized LLMs are more robust. For StarCoder, we discard it on HumanEval dataset since we cannot get any outputs.}

\resizebox{\textwidth}{!}{
\begin{tabular}{l|l|ccc|ccc|ccc|ccc}
\hline
\multirow{2}{*}{Model} & \multirow{2}{*}{Size} & \multicolumn{3}{c|}{MBPP} & \multicolumn{3}{c|}{MBPP$^+$} & \multicolumn{3}{c|}{HumanEval} & \multicolumn{3}{c}{HumanEval$^+$} \\
\cline{3-14}
& & Ch & W & S & Ch & W & S & Ch & W & S & Ch & W & S \\
\hline \hline
\multirow{3}{*}{\textbf{LLaMA}} & 1B & 1.092 $\darkred\Downarrow$ & 1.862 $\darkred\Downarrow$ & 1.333 $\darkred\Downarrow$ & 1.053 $\darkred\Downarrow$ & 1.917 $\darkred\Downarrow$ & 1.900 $\darkred\Downarrow$ & 1.500 $\darkred\Downarrow$ & $\darkgreen\Uparrow$ & 0.714 $\darkgreen\Uparrow$ & 1.333 $\darkred\Downarrow$ & $\darkgreen\Uparrow$ & 0.714 $\darkgreen\Uparrow$ \\
& 3B & 1.136 $\darkred\Downarrow$ & 1.630 $\darkred\Downarrow$ & 0.769 $\darkgreen\Uparrow$ & 1.000  & 1.258 $\darkred\Downarrow$ & 0.559 $\darkgreen\Uparrow$ & 0.625 $\darkgreen\Uparrow$ & $\darkgreen\Uparrow$ & 0.833 $\darkgreen\Uparrow$ & 0.750 $\darkgreen\Uparrow$ & 0.111 $\darkgreen\Uparrow$ & 0.857 $\darkgreen\Uparrow$ \\
& 8B & 0.429 $\darkgreen\Uparrow$ & 0.500 $\darkgreen\Uparrow$ & $\darkgreen\Uparrow$ & 0.621 $\darkgreen\Uparrow$ & 0.594 $\darkgreen\Uparrow$ &  $\darkgreen\Uparrow$ & 1.000 & $\darkgreen\Uparrow$ & 0.786 $\darkgreen\Uparrow$ & 0.400 $\darkgreen\Uparrow$ & $\darkgreen\Uparrow$ & 0.929 $\darkgreen\Uparrow$ \\
\hline
\multirow{3}{*}{\textbf{DeepSeek}} & 1.3B & 1.000  & 0.700 $\darkgreen\Uparrow$ & 1.333 $\darkred\Downarrow$ & 1.219 $\darkred\Downarrow$ & 0.652 $\darkgreen\Uparrow$ & 1.167 $\darkred\Downarrow$ & 0.750 $\darkgreen\Uparrow$ & $\darkred\Downarrow$ & 1.000 & 0.750 $\darkgreen\Uparrow$ & $\darkred\Downarrow$ & 1.250 $\darkred\Downarrow$ \\
& 6.7B & 1.778 $\darkred\Downarrow$ & 1.444 $\darkred\Downarrow$ & 1.133 $\darkred\Downarrow$ & 1.600 $\darkred\Downarrow$ & 1.555 $\darkred\Downarrow$ & 1.571 $\darkred\Downarrow$ & $\darkred\Downarrow$ & $\darkred\Downarrow$ & 5.000 $\darkred\Downarrow$ & 2.000 $\darkred\Downarrow$ & $\darkgreen\Uparrow$ & 4.500 $\darkred\Downarrow$ \\
& 33B & $\darkred\Downarrow$ & 0.718 $\darkgreen\Uparrow$ & 0.488 $\darkgreen\Uparrow$ & $\darkred\Downarrow$ & 0.600 $\darkgreen\Uparrow$ & 0.467 $\darkgreen\Uparrow$ & 0.500 $\darkgreen\Uparrow$ & 0.889 $\darkgreen\Uparrow$ & 1.400 $\darkred\Downarrow$ & 0.429 $\darkgreen\Uparrow$ & 1.167 $\darkred\Downarrow$ & 1.000 \\
\hline
\multirow{3}{*}{\textbf{CodeGen}} & 350M & 0.942 $\darkgreen\Uparrow$ & 0.758 $\darkgreen\Uparrow$ & 0.857 $\darkgreen\Uparrow$ & 1.017 $\darkred\Downarrow$ & 1.000 & 6.997 $\darkred\Downarrow$ & 0.875 $\darkgreen\Uparrow$ & 1.200 $\darkred\Downarrow$ & 1.250 $\darkred\Downarrow$ & 0.833 $\darkgreen\Uparrow$ & 2.000 $\darkred\Downarrow$ & 2.000 $\darkred\Downarrow$ \\
& 2B & 1.029 $\darkred\Downarrow$ & 2.437 $\darkred\Downarrow$ & 1.437 $\darkred\Downarrow$ & 0.829 $\darkgreen\Uparrow$ & 1.661 $\darkred\Downarrow$ & 1.083 $\darkred\Downarrow$ & 0.400 $\darkgreen\Uparrow$ & 3.000 $\darkred\Downarrow$ & 1.667 $\darkred\Downarrow$ & 0.455 $\darkgreen\Uparrow$ & 5.999 $\darkred\Downarrow$ & 1.200 $\darkred\Downarrow$ \\
& 6B & 1.315 $\darkred\Downarrow$ & 1.286 $\darkred\Downarrow$ & 1.667 $\darkred\Downarrow$ & 1.289 $\darkred\Downarrow$ & 1.410 $\darkred\Downarrow$ & 1.571 $\darkred\Downarrow$ & 0.600 $\darkgreen\Uparrow$ & 3.666 $\darkred\Downarrow$ & 0.714 $\darkgreen\Uparrow$ & 0.909 $\darkgreen\Uparrow$ & 2.667 $\darkred\Downarrow$ & 0.833 $\darkgreen\Uparrow$ \\
\hline
\multirow{3}{*}{\textbf{StarCoder}} & 3B & 1.579 $\darkred\Downarrow$ & 1.278 $\darkred\Downarrow$ & 3.999 $\darkred\Downarrow$ & 1.211 $\darkred\Downarrow$ & 1.364 $\darkred\Downarrow$ & 1.000 & -- & -- & -- & -- & -- & -- \\
& 7B & 0.483 $\darkgreen\Uparrow$ & 0.684 $\darkgreen\Uparrow$ & $\darkgreen\Uparrow$ & 0.593 $\darkgreen\Uparrow$ & 1.273 $\darkred\Downarrow$ & $\darkgreen\Uparrow$ & -- & -- & -- & -- & -- & -- \\
& 15B & $\darkred\Downarrow$ & 0.148 $\darkgreen\Uparrow$ & 1.857 $\darkred\Downarrow$ &  $\darkred\Downarrow$ & $\darkgreen\Uparrow$ & $\darkred\Downarrow$ & -- & -- & -- & -- & -- & -- \\
\hline
\end{tabular}}
\label{lab:rq1}
\end{table*}

\subsection{RQ1: Adversarial Robustness Evaluation}

The evaluation results are presented in Table~\ref{lab:rq1}.
We comprehensively compare the robustness of 12 (4 model families $\times$ 3 scales) LLMs pre- and post-quantization under different adversarial attack settings, including character-, word-, and sentence-level on four datasets. We use RRS to measure the robustness between LLMs pre- and post-quantization.
For some special cases\footnote{For example, when applying attack, quantized model gets improvement but original model gets deduction, the RRS would be a negative value.}, we directly use $\darkgreen\Uparrow$ or $\darkred\Downarrow$ without specific values to denote the robustness.
Based on our results, we find that, 51.59\% of cases show that quantized LLMs are more robust than their original versions, 42.86\% of cases show that original LLMs are more robust than their quantization versions, and 5.56\% of experiments show that there are no robustness difference between LLMs pre- and post-quantization. One notable exception is the HumanEval dataset, where the original LLaMA family demonstrates solid robustness, with 15 out of 18 experiments showing greater robustness compared to its quantized version, distinguishing from other model families.

\paragraph{Analysis by attack methods}
\textit{We find that sentence-level attacks generally produce the most pronounced differences in robustness between original and quantized models, with some extreme RRS values, while character-level attacks show more moderate impacts across model families.}
For character-level attacks, we observe mixed results across different model families. Quantized LLMs show better robustness in many cases (\eg LLaMA-1B, DeepSeek-6.7B, and CodeGen-6B), with RRS values greater than 1. However, for larger models like LLaMA-8B and StarCoder-7B, the original versions exhibit better robustness. On HumanEval datasets, most original LLMs present better robustness than their quantized versions, particularly in the LLaMA family. This suggests that the impact of quantization on character-level robustness is both model and dataset-dependent.
For word-level attacks, there is a clear pattern across model scales. Smaller models (1-3B parameters) generally show improved robustness after quantization, as evidenced by LLaMA-1B, LLaMA-3B, and StarCoder-3B. However, larger models ($>$7B parameters) tend to maintain better robustness in their original form, such as LLaMA-8B, StarCoder-15B, and DeepSeek-33B. An exception is the CodeGen family, where quantization improves word-level robustness for 2B and 6B variants on HumanEval, with RRS values as high as 5.999.
Regarding sentence-level attacks, we observe pronounced differences in robustness. The impact varies significantly across model families, with some showing extreme values (\eg StarCoder-3B with RRS=3.999 on MBPP and CodeGen-2B with RRS=5.999 on HumanEval). Overall, we observe that larger models tend to show more stable robustness between their original and quantized versions, while smaller models exhibit more dramatic differences.

\paragraph{Analysis by model families}
The LLaMA family shows a clear mdoel size-dependent pattern in robustness changes on MBPP datasets. The 1B and 3B variants generally become more robust after quantization across all attack types ($\darkred\Downarrow$), while the 8B variant consistently maintains better robustness in its original version ($\darkgreen\Uparrow$). This suggests that the effectiveness of quantization on robustness may have an inverse relationship with model size in the LLaMA family. However, as we mentioned early, original LLaMA models exhibit amazing robustness on HumanEval datasets.

\noindent$\bullet$ \textbf{DeepSeek Family.} They exhibit interesting patterns where the mid-sized 6.7B variant consistently shows better robustness after quantization ($\darkred\Downarrow$). The smallest 1.3B model shows mixed results, with better original robustness for word-level attacks on MBPP datasets and character-level attacks on HumanEval, but improved post-quantization robustness for sentence-level attacks. The larger 33B model maintains better robustness in its original form for word and sentence-level attacks, while showing improved quantized robustness for character-level attacks. This indicates that the relationship between model size and quantization impact on robustness is not strictly linear in the DeepSeek family.

\noindent$\bullet$ \textbf{CodeGen Family}. While not as uniform as initially described, these models show a tendency toward improved robustness after quantization, particularly for word and sentence-level attacks on HumanEval datasets. The 2B and 6B variants demonstrate notably high RRS values for word-level attacks on HumanEval datasets after quantization. However, a clear exception exists for character-level attacks on HumanEval datasets, where all original CodeGen models are more robust than their quantized versions. An exception is from char-level attack on HumanEval datasets, all original CodeGen models are more robust than their quantized versions.

\noindent$\bullet$ \textbf{StarCoder Family} These models present the most varied results, with significant fluctuations in robustness across different sizes and attack types. The 3B variant predominantly shows improved robustness after quantization, while the 7B variant displays better original robustness under character-level attacks and some sentence-level attacks but mixed results for word-level attacks. The 15B variant exhibits inconsistent patterns, with better original robustness for word-level attacks on MBPP but improved post-quantization robustness for character and sentence-level attacks. This variability suggests that StarCoder's robustness characteristics may be more sensitive to both quantization and attack type compared to other model families.
\begin{figure*}[th]
    \centering
    \includegraphics[width=1.0\linewidth]{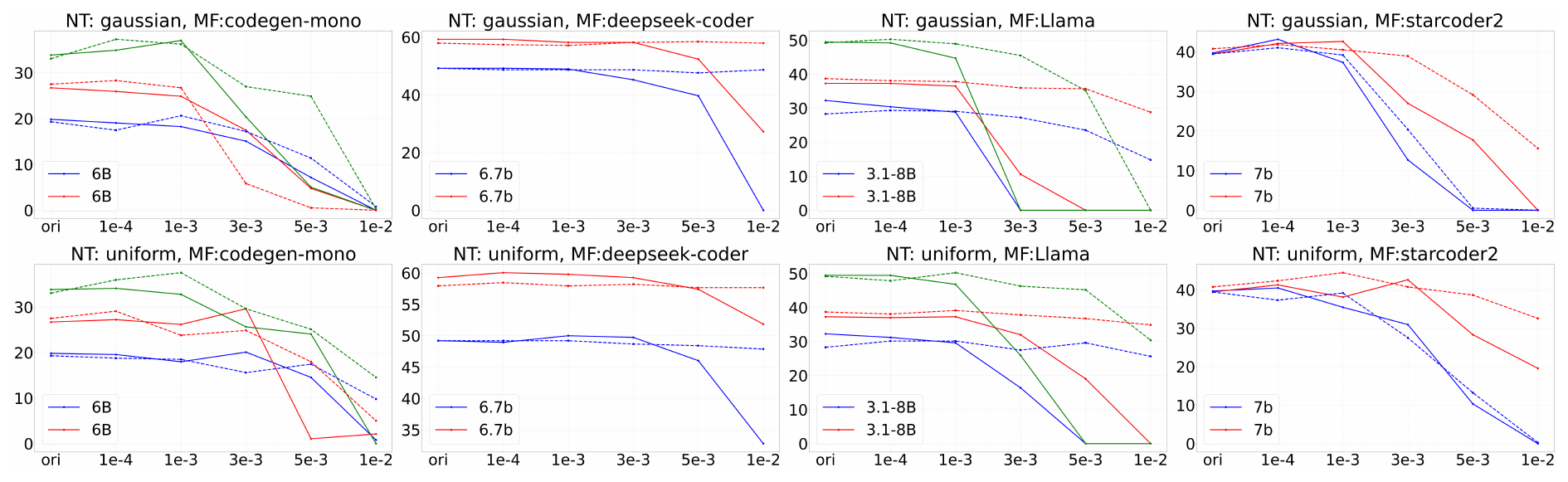}  
    \caption{The results of noise robustness evaluation.}
    \label{lab:rq2}
\end{figure*}
\begin{tcolorbox}[width=\columnwidth]
\textbf{Answer Summary to RQ1:}
\textit{
Quantization's impact on LLM adversarial robustness varies significantly by model size and attack type. Smaller models (1-3B parameters) generally become more robust post-quantization, whereas larger models (>7B) maintain better robustness in their original version; Effects vary by attack type and model family: character-level attacks show model-dependent patterns, word-level attacks reveal size-dependent trends, and sentence-level attacks demonstrate the greatest differences. Model families exhibit distinct patterns: LLaMA shows an inverse size- dependent relationship, DeepSeek displays non-linear patterns, CodeGen generally improves after quantization, and StarCoder shows variable results.
}
\end{tcolorbox}

\subsection{RQ2: Noise Robustness Evaluation}
We evaluate the robustness of LLMs pre- and post-quantization against noise perturbations applied to their model weights. We apply two different types of noise (\eg Gaussian and Uniform) at five increasing intensity levels, \ie 1e-4, 1e-3, 3e-3, 5e-3, and 1e-2 (where higher values indicate stronger noise perturbations), to assess how quantization affects the LLMs' resilience to weight perturbations. The results are presented in Figure~\ref{lab:rq2} and we use pass@1 to measure the performance of each model.

As shown in Figure~\ref{lab:rq2}, we observe a consistent pattern across different LLM families and model sizes: quantized models generally exhibit superior robustness against weight perturbations compared to their full-precision counterparts. Such enhanced robustness is particularly evident when noise levels increase (3e-3 and above), where the performance gap between original and quantized LLMs becomes more pronounced.

\paragraph{Analysis by model families} CodeGen family shows mixed results regarding robustness enhancement through quantization. While CodeGen-350M and CodeGen-6B exhibit improved robustness after quantization, CodeGen-2B presents an interesting exception where the original model demonstrates better resilience to noise than its quantized counterpart. At moderate to high noise levels (3e-3 and above), the original CodeGen-2B maintains higher pass rates than its quantized version. This contrasting behavior within the same model family suggests that the relationship between quantization and robustness may depend on specific architectural characteristics or parameter distributions that vary across model scales, even within the same family.

\noindent$\bullet$ \textbf{DeepSeek family}. We observe that DeepSeek pre- and post-quantization has a similar ability to defend against noise attacks. When the noise level continuously increases, the quantized model could still maintain a relatively high performance while the original model starts to crash. This indicates that quantization improves the range of the noise relisence.

\noindent$\bullet$ \textbf{LLaMA family}. We notice that models display consistent trends across different sizes. For LLaMA-3.2-1B and LLaMA-3.2-3B, the quantized versions consistently outperform their original counterparts at high noise levels (3e-3 and above). For LLaMA-3.1-8B, both the original and quantized models show similar degradation patterns, with performance dropping to nearly zero at the highest noise level (1e-2). However, at the intermediate noise level (5e-3), the quantized LLaMA-3.1-8B maintains significantly better performance compared to its original counterpart. Overall, smaller LLaMA models appear to benefit more consistently from the regularizing effect of quantization than the larger 8B variant.

\noindent$\bullet$ \textbf{StarCoder family}. Our results show a clearly improved robustness through quantization. For both StarCoder-3B and StarCoder-7B variants, the quantized versions consistently maintain higher pass rates at increased noise levels compared to their original counterparts. The difference is particularly pronounced at noise levels 3e-3 and above, where the original models exhibit steep performance degradation while the quantized versions maintain relatively stable performance. For example, at noise level 5e-3, the quantized StarCoder-3B maintains significantly higher performance than its original version.

\paragraph{Analysis by noise types}

\noindent$\bullet$ \textbf{Gaussian noise}. We observe that quantized models generally maintain better performance as noise intensity increases across most model families. In the CodeGen family, quantized versions of CodeGen-350M and CodeGen-6B show superior resilience, while CodeGen-2B exhibits the opposite pattern with the full-precision model performing better. DeepSeek models maintain consistent performance until higher noise levels (5e-3), where quantized variants demonstrate significantly better robustness as the original models' performance drops sharply. For LLaMA models, quantized versions consistently outperform their original counterparts at moderate to high noise levels, though both variants of LLaMA-8B degrade substantially at the highest noise levels. StarCoder shows similar benefits from quantization, with StarCoder-7B displaying the most significant improvement in noise resilience compared to its original version.

\noindent$\bullet$ \textbf{Uniform noise} We observe similar robustness patterns under uniform noise, with quantized models generally demonstrating superior resilience. The CodeGen family shows consistent results with the Gaussian case: quantized versions of CodeGen-350M and CodeGen-6B maintain higher performance at elevated noise levels, while CodeGen-2B shows better robustness in its full-precision form. DeepSeek models exhibit comparable performance between quantized and original variants at low noise levels, but as noise intensity increases to 5e-3 and beyond, the quantized models retain significantly more functionality. LLaMA models follow a similar pattern to their Gaussian noise response, with quantized versions of LLaMA-1B and LLaMA-3B consistently outperforming their original counterparts. StarCoder-7B demonstrates particularly significant benefits from quantization, maintaining substantially higher pass rates even at extreme noise levels, while StarCoder-3B shows minimal difference between its original and quantized versions.

\begin{tcolorbox}[width=\columnwidth]
\textbf{Answer Summary to RQ2:}
\textit{
Quantization generally improves LLMs' robustness against weight perturbations (both in Gaussian and Uniform noise), with the effect becoming more prominent at higher noise levels ($\geq3e-3$). Results vary by model family: CodeGen shows mixed results, DeepSeek maintains performance post-quantization under increasing noise, smaller LLaMA models benefit more from quantization, and StarCoder consistently shows enhanced robustness. These variations suggest quantization's impact on robustness depends on specific architectural characteristics across model scales and families.
}
\end{tcolorbox}

\subsection{Quantization-Robustness Trade-off Evaluation}
\begin{figure*}[t]
    \centering
    \includegraphics[width=1\linewidth]{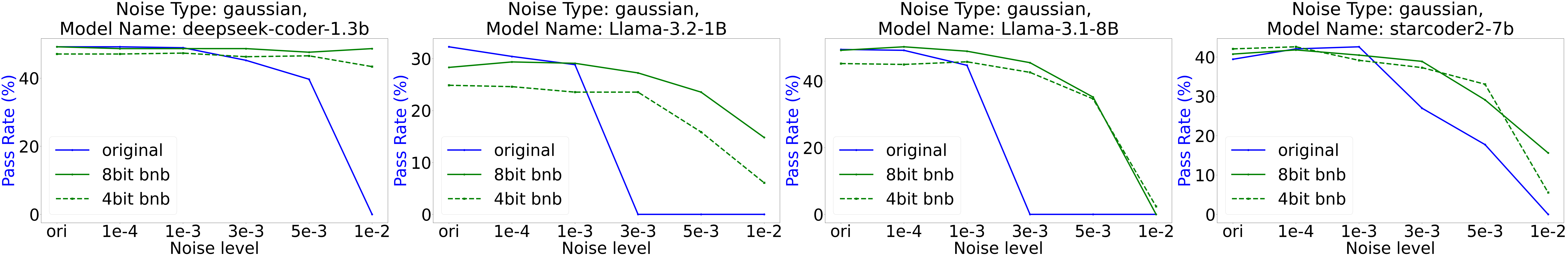}
    \caption{The results of quantization-robustness trade-off evaluation.}
    \label{lab:rq3}
\end{figure*}

\begin{figure}[t]
    \centering
    \includegraphics[width=1\linewidth]{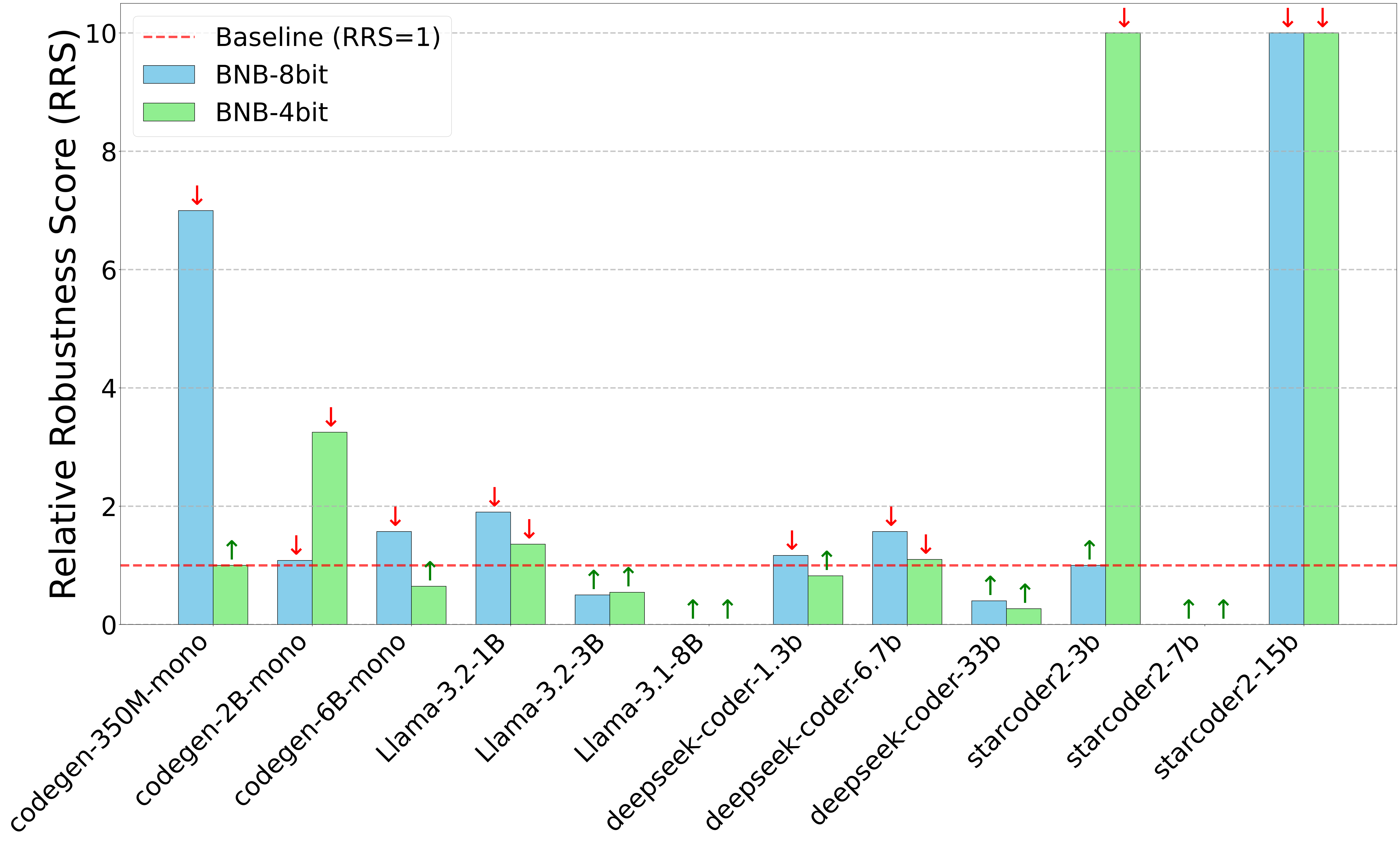}
    \caption{The results of quantization-robustness trade-off evaluation of translation attack.}
    \label{lab:rq3_adv}
\end{figure}

Our investigation reveals a relationship between quantization and robustness in LLMs. We conducted an extensive evaluation campaign spanning 252 experimental configurations (4 model families × 3 sizes × 3 precision levels × (6 noise levels + 1 adversarial attack)) to understand this relationship across different model architectures and sizes. Each model is evaluated at three precision levels (full precision, 8-bit, and 4-bit quantization) while being tested against five increasing levels of Gaussian noise perturbation (1e-4, 1e-3, 3e-3, 5e-3, and 1e-2) plus one adversarial attack (sentence-level), allowing us to identify critical thresholds where performance begins to degrade significantly.

Figure~\ref{lab:rq3} presents a visualization of how different models respond to noise under varying quantization levels. Quantization introduces a modest initial performance penalty—8-bit quantization decreases clean performance by just 0.36\% on average, while 4-bit causes a more noticeable 1.19\% reduction. However, this small sacrifice yields substantial robustness benefits when models encounter noisy inputs. We could observe three distinct regions of behavior: at low noise levels (below 1e-3), all precision variants perform comparably; at moderate noise levels (1e-3 to 5e-3), quantized models begin demonstrating superior resilience; and at high noise levels (5e-3 and above), original full-precision models suffer catastrophic performance collapse while quantized versions—especially 8-bit variants—maintain competitive performance.

Figure~\ref{lab:rq3_adv} presents the differential impacts of 8-bit and 4-bit quantization through Relative Robustness Scores (RRS). The comparison between blue bars (8-bit BNB) and green bars (4-bit BNB) reveals clear patterns across model families. 8-bit quantization delivers superior robustness benefits in several models, including CodeGen-350M, CodeGen-6B, LLaMA-3.2-1B, DeepSeek-1.3B, and DeepSeek-6.7B, where they have a higher RRS (>1). In contrast, 4-bit quantization proves more beneficial for CodeGen-2B and StarCoder-3B. Interestingly, StarCoder-15B shows comparable robustness gains from both quantization levels, suggesting that for some models, the degree of quantization may not significantly impact robustness beyond a certain threshold. Most notably, several models—LLaMA-3.2-3B, LLaMA-3.1-8B, DeepSeek-33B, and StarCoder-7B—actually demonstrate higher robustness in their original versions. These varied responses across model families and sizes highlights that optimal quantization strategies for maximizing both performance and resilience must be determined on a case-by-case basis through empirical evaluation. Generally, smaller LLMs tend to obtain robustness after quantization while larger LLMs has mixed tendency of robustness, which depents specific model architecture.

\begin{tcolorbox}[width=\columnwidth]
\textbf{Answer Summary to RQ3:}
\textit{
Across 252 experimental configurations, quantization could enhance LLM robustness against noise perturbations and adversarial attacks, with model-specific trade-offs. Generally, 8-bit quantization offers optimal balance, reducing performance by only 0.36\% while substantially improving noise resistance. 4-bit quantization provides greater robustness at a higher cost (1.19\% reduction). Smaller models (<3B parameters) handle aggressive quantization better than larger ones, which show architecture-dependent responses. 
}
\end{tcolorbox}

\section{Discussion}

\subsection{Quantization for Robustness?}


\begin{figure*}[th]
\begin{center}
\begin{minipage}[t]{.4\textwidth}
\begin{lstlisting}[
frame=single, 
breaklines=true, 
breakindent=0pt, 
columns=fullflexible, 
mathescape=true,
basicstyle=\footnotesize,
keywords={the, python, Python},
keywordstyle=\color{red},
caption={The prompt from MBPP$^+$ No. 764 task.}, 
captionpos=b, 
label={lst:discussion:exp1}
]
$\textbf{Original prompt:}$ 
"""
Write a python function to count number of digits in a given string.
assert number_ctr('program2bedone') == 1
"""
$\textbf{Adversarial prompt:}$
"""
Write a Python function to count the number of digits in a given string.
assert number_ctr('program2bedone') == 1
"""
\end{lstlisting}
\end{minipage}
\hspace{0.05\textwidth}
\begin{minipage}[t]{.45\textwidth}
\begin{lstlisting}[
frame=single, 
breaklines=true, 
breakindent=0pt, 
columns=fullflexible, 
mathescape=true, 
basicstyle=\footnotesize,
keywords={given, specified, any, no, none},
keywordstyle=\color{red},
caption={The prompt from MBPP$^+$ No. 744 task.}, 
captionpos=b, 
label={lst:discussion:exp2}
]
$\textbf{Original prompt:}$ 
"""
Write a function to check if the given tuple has any none value or not.
assert check_none((10, 4, 5, 6, None)) == True
"""
$\textbf{Adversarial prompt:}$
"""
Write a function to check if the specified tuple has no value or not.
assert check_none((10, 4, 5, 6, None)) == True
"""
\end{lstlisting}
\end{minipage}
\end{center}
\caption{The examples in MBPP$^+$ before/after sentence-level adversarial attack. }
\label{fig:disccusion}
\end{figure*}

\noindent\textit{\textbf{Quantization as noise.}} 
Quantization introduces controlled noise into model parameters, which can be regarded as an implicit form of regularization. By reducing the precision of weights, quantization constrains the parameter space, limiting the model's ability to overfit to non-robust features in the data. Such a regularization encourages LLMs to rely on more robust features that are better for generalization under adversarial conditions.
As shown in Listing~\ref{lst:discussion:exp1}, our adversarial attack only introduce slight perturbation, transforming ``p'' in python into ``P'' and adding a word ``the'' to the front of the word ``number'', but original LLMs cannot generate correct output for it while their quantized versions could do that.
Our experimental results align with the observations from \citeauthor{tsipras2018robustness}, who show that robust models learn fundamentally different features than standard models, and quantization could potentially be regarded as robustness features~\cite{tsipras2018robustness}. \\
\noindent\textit{\textbf{Reduced parameter sensitivity.}} Quantization potentially reduces parameter sensitivity through its effect on feature representation. When model weights are quantized to lower bit precision, the model is forced to represent information more efficiently, focusing on the most essential features in code while discarding subtle or spurious correlations that full-precision models might overfit to. Such a feature simplification process naturally reduces sensitivity to parameter perturbations. With fewer bits to represent weights, quantized models develop more distinct and robust decision boundaries that are less affected by small adversarial shifts. Our noise perturbation experiments demonstrate this phenomenon clearly—quantized LLMs withstand higher levels of weight disturbances. The discrete nature of quantized weights creates natural thresholds to shield model behavior, whereas full-precision models can be influenced by infinitesimal changes to their parameters. As shown in Listing~\ref{lst:discussion:exp2}, when we replace some words in original prompt, it actually introduces slight semantical ambiguity. However, original LLMs are sensitive to this change and generate wrong solution, while their quantized versions could still generate accurate solutions by mitigating the ambiguity with the reference of the given test case example.

\subsection{Implications}


Quantization can serve as a dual-benefit strategy that not only reduces computational requirements but also enhances model robustness, particularly for small to medium-sized models (1-7B parameters). Generally, our findings suggest size-dependent quantization recommendations: \textbf{smaller models (350M-3B) benefit from quantization with minimal performance impact; medium-sized models (3B-7B) achieve optimal balance with 8-bit quantization; and larger models (>7B) require careful consideration as effects vary by model family.} For application-specific strategies, \textbf{general-purpose code generation benefits the most from 8-bit quantization and resource-constrained deployments could consider aggressive quantization of smaller models.} Different model families show distinct patterns: CodeGen exhibits consistent improvement, LLaMA shows size-dependent benefits, DeepSeek displays non-linear improvement, and StarCoder presents variable results in different sizes. These insights can guide in deploying LLMs that are both efficient and reliable for code generation tasks.

\subsection{Threats to Validity}
Our study on quantized LLMs' robustness in code generation faces several potential threats to validity that warrant acknowledgment:

\noindent\textit{\textbf{Threats to internal validity.}} The observed robustness improvements might be influenced by factors beyond the quantization itself, such as our chosen quantization method, which could affect results differently than alternative quantization techniques. Additionally, our implementation choices for the three adversarial attacks may impact results: the specific parameter settings such as the character-level attack's specific transformation parameters (0.5 probability, 5-character maximum). To mitigate these internal threats, we maintained consistent hyperparameters across all experiments for both original and quantized models, verified semantic preservation through manual inspection of attack outputs, calibrated parameters to ensure comparable perturbation intensities across attack levels, and designed our RRS metric specifically to normalize performance changes and enable fair comparisons between original and quantized models, reducing the impact of baseline performance differences.

\noindent\textit{\textbf{Threats to construct validity.}} Discrepancies between theoretical expectations and observed outcomes can impact the validity of our investigation. A primary concern in this context is the selection of metrics used to evaluate the robustness of LMCs, as well as the methodologies employed to implement adversarial attacks. To mitigate these risk, we rely on well-established metrics and state-of-the-art approaches to attack LLMs.
In particular, we implement multiple types of adversarial attacks (character-level, word-level, and sentence-level) and noise perturbations (Gaussian and Uniform) to ensure our robustness assessments were comprehensive across different perturbation types. While domain limitations remain, these measures help to keep the broader applicability of our findings within the code generation domain.


\section{Conclusion}
In this paper, we conducted the first comprehensive investigation of how quantization affects the robustness of LLMs in code generation tasks. Through extensive experiments across four prominent LLM families with multiple scales, we analyzed the robustness from dual perspectives: adversarial attacks on input prompts and noise perturbations on model weights. Our findings challenge the conventional wisdom that quantization necessarily involves a trade-off between efficiency and model quality. We discovered that quantized LLMs frequently demonstrate superior robustness compared to their full-precision counterparts, with 59.72\% of our adversarial experiments showing better resilience in quantized models. This pattern was particularly pronounced in smaller to medium-sized models (1-7B parameters). Besides, quantization could consistently enhance robustness across different attack methods. Similarly, our noise perturbation experiments confirmed that quantized models could generally withstand higher levels of weight disturbances, suggesting that quantization not only reduces computational requirements but can actually improve model reliability.

In the future, we plan to extend this investigation with additional programming languages, model architectures, and quantization techniques. We also plan to explore more theoretical research to fully understand the mechanisms by which quantization enhances model robustness, prompting advanced quantization techniques specifically designed to optimize for both efficiency and robustness.

\balance
\bibliography{IEEEabrv,ref}
\end{document}